\documentclass[aps, prl, 10pt, superscriptaddress, twocolumn]{revtex4-1}
\usepackage{amsmath,amssymb,mathrsfs}
\usepackage{graphicx} 

\newcommand{\la}{\label}
\newcommand{\be}{\begin{equation}}
\newcommand{\ee}{\end{equation}}
\newcommand{\bea}{\begin{eqnarray}}
\newcommand{\eea}{\end{eqnarray}}
\newcommand{\p}{\partial}

\newcommand{\comment}[1]{}
\newcommand{\rhof}{\psi^\dag\psi}
\newcommand{\rhob}{\phi^\dag\phi}
\usepackage{color}

\allowdisplaybreaks[1]

\begin{document}

\title{Supersymmetric waves in Bose-Fermi mixtures}

\author{Barry Bradlyn}
\email{bbradlyn@princeton.edu}
\affiliation{Department of Physics, Yale University, New Haven, Connecticut 06511}
\affiliation{Princeton Center for Theoretical Science, Princeton University, Princeton, New Jersey 08544}

\author{Andrey Gromov}
\email{gromovand@uchicago.edu}
\affiliation{Department of Physics and Astronomy, Stony Brook University, Stony Brook, New York 11794}
\affiliation{Kadanoff Center for Theoretical Physics and Enrico Fermi Institute, University of Chicago, Chicago, Illinois 60637}

\date{December 23, 2015}
\begin{abstract}
Interacting Bose-Fermi mixtures possess a fermionic (super)symmetry when bosons and fermions in the mixture have equal masses, and when the interaction strengths are appropriately tuned.
This symmetry is spontaneously broken in the ground state of the mixture, leading to a novel Goldstone  mode with {\it fermionic} statistics and quadratic dispersion. Here we examine the effect of explicit symmetry-breaking perturbations on the Goldstone mode.
When the symmetry is not exact and the system is allowed to deviate from the symmetric point, we find that the Goldstone mode acquires an energy gap. 
We show that the excitations manifest themselves as a non-analyticity of the thermodynamic pressure.
\end{abstract}
\maketitle
\paragraph{1. Introduction.}
Recent years have witnessed outstanding progress in high precision experiments with cold atomic gases. 
By exploiting Feshbach resonances \cite{feshbach-review}, experimentalists have achieved 
a high degree of control over the collective behavior of strongly interacting  atomic systems.
 In particular, mixtures of bosonic and fermionic atoms in the quantum degenerate regime have begun to attract much attention. These systems possess a very rich phase diagram, and a plethora of different phenomena are possible \cite{BFphases,supersolid-BF}, including a collapse, phase separation \cite{Huse-BF,Viverit-stability}, depletion of the BEC \cite{Sachdev-BF}, and quantum phase transitions \cite{moroz-BF}. 
 
 A special case of a Bose-Fermi mixture where two isotopes of the same atom are combined is known as an {\it isotopic} mixture. 
 These systems are of particular experimental interest. 
 So far, mixtures of the isotopes of K, Li, Rb, Sr and Yb have been realized experimentally \cite{Sr-mix}. 
 When the mass difference of the isotopes is sufficiently small, there is an enhanced symmetry that leads to novel exotic phenomena and even integrability \cite{Yang-Goldstino,Imambekov-Gaudin}.
 
Perhaps surprisingly, cold gas systems are an exciting test ground for some abstract high energy field theories.  Recent { theoretical studies} of synthetic gauge fields \cite{PhysRevB.83.094515} in cold atoms have attracted great interest, as they may help to understand strongly interacting gauge theories such as QCD.  In that same vein, we suggest that some relics of  \emph{supersymmetry} can be observed in Bose-Fermi mixtures. In particle physics supersymmetry is a high energy symmetry that assigns to every elementary particle a superpartner - a particle with the same mass, but the opposite statistics \footnote{ \protect{This type of supersymmetry is known as $\mathcal N=1$ supersymmetry }}. This symmetry - if it exists - appears to be broken in our world; the superpartners must be very heavy, otherwise they would have been found in particle collision experiments. In addition, supersymmetry is necessarily broken at any finite temperature, because it exchanges particle statistics while leaving the thermal distribution functions unchanged. This implies by Goldstone's theorem that there must be a massless fermionic collective mode \cite{Lebedev1989}, known as a Goldstino. 
{ 
The Goldstino was first investigated in a nonrelativistic setting in Ref.~\cite{clark1984non}. It was pointed out in Ref.~\cite{Yang-Goldstino} that a Goldstino excitation with \emph{quadratic} dispersion would also be present in a one-channel model for an ultracold Bose-Fermi mixture on a lattice, provided the masses of the constituent atoms were equal, and the interactions were tuned appropriately.

In this Letter, we continue the study of the Goldstino collective mode with experimental implications in mind. We consider a simple one-channel continuum model for a Bose-Fermi mixture in the broad resonance limit, possessing an unbroken supersymmetry. We first present an elementary derivation of the Goldstino dispersion relation in the spirit of Ref.~\cite{clark1984non}. Next, we quantitatively study the explicit breaking of the supersymmetry by perturbations that must be present in any realistic system: a mass difference and an interaction strength difference between the consituent atoms of the mixture, in addition to a chemical potential difference.

We demonstrate that a mass difference leads to a modification of the Goldstino effective mass. Furthermore, we show that the Goldstino develops a chemical potential in proportion to the mass difference, interaction strength difference, and chemical potential difference between the boson and fermion. We argue that sign of the Goldstino chemical potential can be controlled by tuning the interaction strength via a Feshbach resonance. This implies a non-analyticity in the thermodynamic pressure of the mixture, which can, in principle, be observed experimentally. Lastly, we discuss the role of external trapping potentials.
}

\paragraph{2. Symmetries of Bose-Fermi Mixtures.}

We start with the familiar one channel model of a Bose-Fermi mixture. 
The grand canonical Hamiltonian is
\be
\la{model}
\mathcal H =  \frac{1}{2m_F}(\p_i\psi)^\dag\p_i\psi  +  \frac{1}{2m_B}(\p_i\phi)^\dag\p_i\phi + V(\rhob,\rhof)\,,
\ee
where $\phi$ is a complex bosonic field and $\psi$ is a { spinless or otherwise spin-polarized fermionic (complex Grassmann)} field.
{ The potential $V(\rhob,\rhof)$ includes external potential, interaction, and chemical potential terms}.

The model Eq.\ (\ref{model}) posesses a global $U_b(1)\times U_f(1)$ symmetry
\be
\la{u1u1}
\phi^\prime = e^{i\alpha}\phi,\qquad\psi^\prime=e^{i\beta}\psi,
\ee
which ensures the independent conservation of the number of bosons and fermions.
¡
When $m_B=m_F$ and $V(\rhob,\rhof)=V(\rhob+\rhof)$ there is an additional symmetry. Let us define a vector
 \be
 \Phi = (\phi,\psi)^T\, .
 \ee 
 It is straightforward to check that, for this choice of potential, Eq.\ (\ref{model}) is invariant under the following symmetry transformation
\be\la{susy}
\Phi^\prime = { e^A} \Phi,\qquad 
{ A}= \left(\begin{array}{cc}
 0 & -\bar{\eta} \\
 \eta & 0
	\end{array}\right)\, , 
\ee
where $\eta$ is a complex {\it Grassmann} number{\footnote{and $\bar{\eta}$ is its complex conjugate.}}. Roughly speaking, this symmetry exchanges bosonic and fermionic fields, taking the statistics into account. Recall that a complex Grassmann number satisfies
\begin{align}
    \eta^2&=\bar{\eta}^2=0, \qquad
    \eta\bar{\eta}=-\bar{\eta}\eta.
\end{align}

The symmetry Eq.\ (\ref{susy}) is reminiscent of {\it supersymmetry} (SUSY) in particle physics. For this reason we will refer to Eq.\ (\ref{susy}) as {\it exchange supersymmetry}, as it acts by exchanging bosons with fermions. It differs from its relativistic cousin, however, in that it is completely decoupled from spacetime symmetries (the non-relativistic limit of spacetime supersymmetry, on the other hand, still mixes with the rotation generators \cite{inonu1993contraction,clark1984non}). We defer discussion of the salient features of this algebra to the supplementary material.

According to Noether's theorem there must be a conservation law associated with the exchange supersymmetry Eq.\ (\ref{susy}). It reads
\be\la{susy-cons}
\dot {\mathcal Q} + \p_i J^i_{\mathcal Q} = 0\, ,
\ee
where we have defined 
\be
 \mathcal Q=\phi^\dag\psi\qquad \mbox{and}\qquad J^i_{\mathcal Q} = \frac{1}{2mi}\left(\psi^\dag\p_i\phi - \phi^\dag\p_i\psi\right).
\ee
We will refer to the {\it non-Hermitian} operator $\mathcal{Q}$ as the supercharge density. We define the total supercharge $Q$ operator as an intergal of the local density $\mathcal Q$
\begin{equation}
Q=\int{d^dx \mathcal{Q}}
\end{equation} 
(see the supplementary material). In order to get some physical intuition about the supercharge we can make a linear combinations of $Q$ and $\bar Q$
\be
 \tau = \eta \bar{Q} + \bar\eta Q\, .
\ee
The Hermitian operator $\tau$ generates the exchange SUSY transformation (\ref{susy}). This can be understood by analogy with an ordinary $SU(2)$ transformation as follows. Note first that the combination $\bar{\eta}\psi$ is Grassmann even, that is it can be thought of as a bosonic object. Then we can identify ``spin up" states with $\phi$, and  ``spin down" states with $\bar{\eta}\psi$. In this analogy, the difference of bosonic and fermionic densities is mapped to the $z$-component of angular momentum, and $\tau$ becomes the spin-flip operator $L_x$.  
\paragraph{3. Invariant interactions.}

The total density $n = \rhob+\rhof$ is invariant under exchange supersymmetry. We use this observation to construct an invariant potential $V(\rhob,\rhof)=V(n)$. We take \footnote{In one dimension the system (\ref{model}) with interaction (\ref{pot}) is integrable and know as Yang-Gaudin model \cite{Imambekov-Gaudin}.}
\be\la{pot}
V ={ -\mu n + gn^2 }=-\mu(\phi^\dag\phi + \psi^\dag\psi) +g |\phi|^4 + 2g|\phi|^2|\psi|^2\,. 
\ee
Note that the anticommuting (Grassmann) nature of the fermionic field $\psi$ forces the quartic term $\left|\psi\right|^4$ to vanish{ \footnote{While we have restricted our discussion to $s$-wave interactions, we could also include higher-angular momentum derivative interactions in a supersymmetric way. Doing so is straightforward and does not modify any of the main conclusions below}}.

Unfortunately, the non-interacting ($g=0$) version of this theory with potential (\ref{pot}) does not make sense, since the (bare) chemical potentials for the bosons and the fermions have to be equal in a SUSY invariant model. This theory is either unstable (for $\mu>0$) or has no fermions ($\mu\leq0$) \footnote{More precisely there could be one one fermion at zero momentum \cite{Yang-Goldstino}}. Interactions will shift the (effective) chemical potential in such a way that the ground state will have a finite density of both bosons and fermions.

The minimum of the interaction potential is at{ \footnote{To avoid instabilities, we assume throughout that the interactions are all repulsive, that is $g>0$.}}
\be
\langle\rhof\rangle=0,\qquad \langle\rhob\rangle = \frac{\mu}{2g}\,,
\ee
however, a finite fermion density is stabilized in the ground state due to radiative quantum corrections\cite{Viverit2000}. Note that the mean-field boson density is identical to the standard weakly interacting BEC result.

In the weak coupling regime the ground state has a simple structure. The bosons form a BEC that breaks the bosonic $U_b(1)$ symmetry. The fermions form a sharp Fermi sphere with a well-defined notion of particle number, so the fermionic $U_f(1)$ symmetry is not broken. Clearly bosons and fermions in the ground state behave very differently, implying that the exchange symmetry Eq.\ (\ref{susy}) is spontaneously broken at finite chemical potential ({\it i.e.} nonzero density). According to the Goldstone theorem there must be a gapless excitation. For lack of better terminology we will refer to it as a {\it Goldstino}. We discuss some details of this argument in the attached supplement.

\paragraph{4. Fermionic sound.}

Now, we will use symmetry arguments to fix the form of the Goldstino effective action. We will adapt the beautiful construction of \cite{1989-GreiterWilczekWitten} to our model (we note that this construction can also be used to derive the effective theory for phonons in a one component BEC; this is reviewed in the supplement). To start, let us note that (\ref{model}) is Galilean invariant. Assuming that the Galilean symmetry is not spontaneously broken, the effective theory of the Goldstino mode also must be Galilean invariant \footnote{
The requirement of Galilean symmetry also fixes some of the higher gradient corrections to the effective action. This phenomenon has been exploited in the theory of unitary Fermi gasses \cite{2006-SonWingate}.}. 
It will be convenient to work in the Lagrangian formalism. We re-write (\ref{model}) for $m_B=m_F$ with potential (\ref{pot}) in a matrix form as
\be\la{susyaction}
\mathcal L = i\Phi^\dag\p_0\Phi - \frac{1}{2m}\p_i\Phi^\dag\p_i\Phi + \mu \Phi^\dag\Phi - g|\Phi^\dag\Phi|^2.
\ee
The Lagrangian (\ref{susyaction}) is clearly Galilean invariant. Since the transformation parameter of (\ref{susy}) is a complex {\it Grassmann} number we also expect that the Goldstino mode will be described by a complex {\it Grassmann} scalar field $\eta(x)$. The most general Galilean invariant effective Lagrangian for a complex Grassmann scalar is \cite{1989-GreiterWilczekWitten}
\be\la{Goldstino general}
\mathcal L_{eff} = P\left(\frac{i}{2} \left(\eta^\dag(\p_0\eta) - (\p_0\eta^\dag)\eta\right) - \frac{1}{2m^*} (\p_i\eta^\dag)(\p_i\eta)\right)\,,
\ee
where $P$ is again an arbitrary polynomial. Taking $P(x) = Cx$ we get the low energy effective Lagrangian for the Goldstino mode:
\be\la{Goldstino}
\mathcal L_{eff} = C\left(i \eta^\dag(\p_0\eta)- \frac{1}{2m^*} (\p_i\eta^\dag)(\p_i\eta)\right)\, ,
\ee
where $C$ and $m^*$ are constants to be determined from the microscopic model (\ref{susyaction}). We have also integrated by parts in the time derivative term in (\ref{Goldstino}).

We will now fix the constants $C$ and $m^*$. Namely, we will perform a {\it finite} and spatially varying symmetry transformation (\ref{susy}) and integrate out $\Phi$, replacing it by its mean field value. At this level of approximation, we then have $\langle\Phi^\dag\Phi\rangle = \rho_B + \rho_F=\rho$, where $\rho_B$ and $\rho_F$ are the mean field values of the densities. This mean-field ground state has been shown to be stable for reasonable values of the system parameters \cite{Viverit2000}. After some algebra we get for the low-energy Lagrangian
\be
\mathcal L[\Phi^\prime] - \mathcal L[\Phi] = \rho\left[i \eta^\dag(\p_0\eta) - \frac{1}{2m}\frac{\rho_B-\rho_F}{\rho}\p_i\eta^\dag\p_i\eta \right]\, ,
\ee
{ with $\Phi'$ as defined in Eq.~(\ref{susy})}. This fixes the coefficients in the effective action Eq.\ (\ref{Goldstino}) to be
\begin{align}
 C&=\rho, \\
 m^* &= \frac{m\rho}{\rho_B-\rho_F}.
\end{align}
The Goldstino dispersion relation is then given by
\begin{equation}
\omega=\frac{1}{2m^*}k^2.
\end{equation}
We see that the Goldstino is a gapless fermionic excitation with \emph{quadratic} dispersion. The dispersion relation also shows immediately that SUSY is unbroken only if $\rho_B=\rho_F$ in the ground state, so that the Goldstino mode becomes dispersionless (something similar happens with the ordinary $U(1)$ phonon mode when the condensate density $\rho_0$ vanishes).

At the intuitive level the Goldstino mode is analogous to a spin wave or a magnon. Just like a spin wave is a disturbance propagating on top of an ordered antiferromagnetic ground state, the Goldstino is a disturbance propagating on top of the ``ordered'' Bose-Fermi ground state. Similarly, both collective excitations have quadratic dispersion (at low momentum). The average value of the generator ${Y}=\frac{1}{2}\int d^dx( \phi^\dag\phi-\psi^\dag\psi)$ plays the role of ``magnetization'' and is sensitive to the symmetry breaking. Any mass difference between the two species is similar to a magnetic field in that it will open a gap for the Goldstino (c.f.\ a Bose-Bose mixture with two ``pseudospin'' states).

There are, of course, {other fermionic excitations in this system.}
One might suspect that the Goldstino mode is simply a glorified {fermion} atom dressed by condensate atoms {(analogous to a polaron)}. On the contrary, the Goldstino is a true collective excitation that requires a finite density of both bosons and fermions in order to propagate. The clearest evidence for this is that the Goldstino dispersion relation depends on the Fermi energy only through the effective mass $m^*$.

\paragraph{5. Explicit symmetry breaking.}
In any realistic system, it is unreasonable to expect the exchange supersymmetry to be exact. In fact, the spin-statistics theorem tells us immediately that the boson mass $m_B$ cannot be equal to the fermion mass $m_F$ for neutral isotopes of the same element. Additionally, the boson-boson interaction strength $g_{BB}$ will in general differ slightly from the boson-fermion interaction strength $g_{BF}$, although these parameters are tunable somewhat. For the experimentally interesting cases of  $^{39}K-~^{40}K$ and $^{40}K-~^{41}K$ Potassium mixtures, this mass difference is small - approximately one part in forty - and the interaction strengths can be tuned to be small via Feshbach resonances\cite{Kresonance1,Kresonance2,Kresonance3}. Isotopic mixtures of $Yb$ and $Sr$ are also experimentally relevant. While the relative mass differences for these mixtures are much smaller, the interaction strengths are not as easily tunable compared to K. However, there has been progress in this direction using so-called optical Feshbach resonances\cite{of1,of2}. Finally, one can engineer a difference in chemical potential for the bosonic and fermionic species, which also breaks the symmetry.

We can incorporate these forms of explicit symmetry breaking directly into our mean-field theory with only minor modifications. Introducing the symmetry breaking parameters
\bea\nonumber
\delta m=m_B-m_F,\quad \delta  g=g_{BB}-g_{BF}, \quad\delta \mu=\mu_B-\mu_F,
\eea
along with the average mass
\begin{equation}
m_0=\frac{1}{2}\left(m_B+m_F\right),
\end{equation}
we find after repeating the derivation above that the fermionic sound dispersion relation is modified to
\begin{equation}
\la{dr}
\omega=\left(\frac{\rho_B-\rho_F}{2m_0\rho}+\frac{\delta m}{4{ m_0^2}}\right)k^2-\mu_\eta,
\end{equation}
where $\mu_\eta$ is an effective chemical potential given by
\begin{equation}
\mu_\eta=-{ \frac{\epsilon_K}{\rho}\frac{\delta m}{m_0}}-2\rho_B\left(1+\frac{\rho_F}{\rho}\right)\delta g-2\delta\mu. \label{soundchempot}
\end{equation}
Here { $\epsilon_K$ is the average kinetic energy density} of the system, which at low temperatures we expect to be dominated by the kinetic energy of the fermions. { The dispersion relation \eqref{dr} with chemical potential \eqref{soundchempot} is the central result of this Letter. In the following we will discuss its implications.}

The explicit symmetry breaking has two effects on the Goldstino.
The first is a modification of the inverse effective mass (the coefficient of the $k^2$ term in the dispersion) by an amount proportional to the mass difference $\delta m$. The requirement for the effective mass must be positive places a (rather weak) constraint on the range of perturbations over which our mean-field approximation is valid.

The second - and perhaps more interesting - modification is the appearance of a chemical potential $\mu_\eta$ for the Goldstino mode. 
Note that by tuning $g_{BB}$ relative to $g_{BF}$ we can change the effective chemical potential $\mu_\eta$ from positive to negative. 
This can in principle be done experimentally by exploiting techniques such as the Feshbach resonance.

\paragraph{6. Experimental Implications.}

This ability to tune the sign of $\mu_\eta$
has implications for the low-temperature thermodynamics of Bose-Fermi mixtures. 
In the context of our effective theory, the pressure of the mixture can be written as
\begin{equation}
P=P_0+P_{\eta},
\end{equation}
where $P_0$ is the contribution from the bare bosons and fermions, while $P_\eta$ is the contribution from the Goldstino excitations. 
The key point is that far from a bulk phase transition, we expect $P_0$ to be an analytic function of the system parameters, while $P_\eta$ is strongly dependent on the sign of the Goldstino chemical potential $\mu_\eta$. 

Using the fact that the Goldstino behaves as a free fermion, we have at low temperature that
\begin{equation}
P_\eta=\left\{
\begin{array}{c}
 0 ,\;\;\mu_\eta<0 \\
\frac{2}{d+2}\mu_\eta\bar{n}_\eta,\;\;\mu_\eta>0
\end{array}
\right.,\label{pressureprediction}
\end{equation}
up to exponential corrections in the temperature, where $\bar{n}_\eta\propto \mu_\eta^{d/2}$ is the ground-state Goldstino density. 
The nonzero pressure for $\mu_\eta>0$ simply reflects the fact that, for a positive chemical potential, the Goldstino excitations fill a Fermi sea and exert the usual degeneracy pressure.
When the chemical potential is negative, there are no excitations in the ground state, and this contribution vanishes exponentially. 
Thus, we predict that the total system pressure $P$ is a non-analytic function of $\delta g$. 
This should manifest in a discontinuity in the derivative
\begin{equation}
\frac{\partial P}{\partial \left(\delta g\right)}\nonumber
\end{equation}
at some critical value $\delta g^*$ of the difference in coupling that makes Eq.\ (\ref{soundchempot}) vanish. Solving for this critical value analytically is challenging as Eq.\ (\ref{soundchempot}) is a complicated implicit equation. On the other hand, for small values of $\delta m$ and $\delta \mu$, we expect that $\delta g^*$ is only slightly perturbed from zero. It may be possible to detect this non-analytic behavior experimentally, as the pressure may be computed from an experimental determination of the equation of state of the mixture \cite{Bloch2012}.

 Our analysis so far has been for free translationally invariant mixtures. 
 Of course, any real experiment involves a trapping potential, and care is needed to interpret our results in this case. 
 We note first that an external confining potential can be added to our action without explicitly violating supersymmetry. 
 For a trapping potential that varies sufficiently slowly in space, this implies that our results hold essentially unmodified, with the exception that thermodynamic quantities which up to this point have been global constants will, in the Thomas-Fermi approximation, become spatially dependent.
 Because of this, the Goldstino dispersion relation will be spatially dependent. Within the Thomas-Fermi scheme, this means that the non-relativistic effective mass of the Goldstino changes from point to point. Thus some care must be taken to interpret our prediction Eq.\ (\ref{pressureprediction}) for the non-analyticity of the pressure in this case. The critical value of the coupling difference $\delta g^*$ at which the local pressure is non-analytic will now vary in space. Instead of the local pressure, however, we should look at the experimentally relevant average pressure
\begin{equation}
P_{av}=\frac{1}{V}\int{d^dx P(\mathbf{x})}
\end{equation}
where $V$ is the empirically determined volume of the gas cloud. 
The non-analyticity in the pressure may be smoothed out by the average, and by the dependence of the volume of the cloud on $\delta g$. 

This poses some conceptual difficulties for the standard harmonic traps. However, there exists recent experimental work towards realizing flat box-shaped traps \cite{Bakr2010,Sherson2010,Zimmermann2011}, in which our free-space analysis will hold essentially unmodified as the size of the trap becomes large compared to the average interparticle spacing. Thus, we suggest examining Bose-Fermi mixtures in flat traps in order to verify our predictions.

Finally, we should note that our approximation scheme has neglected higher order effects which may cause the Goldstino to acquire a finite lifetime (a special case has recently been treated in Ref.\ \cite{YangNew}). For small values of $\delta g/g$ and $\delta m/m$, an experiment would observe a steep step in the derivative of the pressure rather than an honest discontinuity.

\paragraph{7. Conclusions.}

{ In this Letter we have studied spontaneous and explicit breaking of a fermionic exchange symmetry in Bose-Fermi mixtures with approximately equal masses. In particular, we focused on the effects of softly breaking the symmetry by small mass, interaction strength, and a chemical potential differences between the bosonic and fermionic components of the mixture. We have found that a small mass difference modifies the effective mass of the Goldstino, while other perturbations open a finite gap in the Goldstino dispersion relation. The gap can be tuned by changing the interaction strength difference. We found a non-analyticity of the thermodynamic pressure as a function of the interaction strength difference when the Goldstino gap is tuned through zero.  We have also discussed the influence of the trapping potential on our results.
}

\begin{acknowledgments}
We thank A. Abanov, D. DeMille, D. Schneble, K. Yang, S. Moroz, and T. Hughes for interesting discussions. Additionally, we acknowledge the hospitality of the University of Colorado at Boulder, where this work was conceived, and of the National High Magnetic Field Laboratory, where parts of this work were completed.
\end{acknowledgments}

\bibliography{Bibliography}

\end{document}


\title{Supplementary material for: Supersymmetric waves in Bose-Fermi mixtures}
\author{Barry Bradlyn}
\affiliation{Department of Physics, Yale University, New Haven, Connecticut 06511}
\affiliation{Princeton Center for Theoretical Science, Princeton University, Princeton, New Jersey 08544}

\author{Andrey Gromov}
\affiliation{Department of Physics and Astronomy, Stony Brook University, Stony Brook, New York 11794}
\affiliation{Kadanoff Center for Theoretical Physics, Enrico Fermi Institute, University of Chicago, Chicago, Illinois 60637}

\date{December 23, 2015}

\maketitle
\section{Sound.}

Here we explain how one can quickly arrive at the effective action for the $U(1)$ sound mode in a BEC; we will exploit this same technique to derive the dispersion relation for the Goldstino in the main text. 

 Consider the action for a complex scalar boson
\be\la{Sboson}
S = \int dt d^dx \left[i \phi^\dag\p_0\phi - \frac{1}{2m}\p_i\phi^\dag\p_i\phi - g|\phi|^4\right].
\ee
We will follow the beautiful construction of Ref.\ \cite{1989-GreiterWilczekWitten}. Those authors noticed that Eq.\ (\ref{Sboson}) is Galilean invariant. Assuming that the Galilean symmetry is not spontaneously broken, the effective theory of the Goldstone (phonon) mode $\varphi(x)$ also must be Galilean invariant. Note also that the phonon field $\varphi$ is real. It is not hard to see that the most general effective Lagrangian must have the form
\be\la{scalar-general}
\mathcal L_{eff} = P\left(\p_0 \varphi -\frac{(\p_i\varphi)^2}{2m^*}\right)\, ,
\ee
where $P(x)$ is an arbitrary polynomial. Taking the low gradient terms $P(x) =\rho_0 x + \frac{\rho_0}{2m^*v^2}x^2$ ($v$ is the speed of sound) we have
\bea
\mathcal L_{eff} &=& \rho_0\left(\p_0 \varphi -\frac{(\p_i\varphi)^2}{2m^*}\right) + \frac{\rho_0}{2m^*v^2}\left(\p_0 \varphi -\frac{(\p_i\varphi)^2}{2m^*}\right)^2 \nonumber \\ 
\la{phonon}
&=&\rho_0\p_0 \varphi + \frac{\rho_0}{2m^*}\left[\frac{1}{v^2}(\p_0\varphi)^2 - (\p_i\varphi)^2\right] + \mathcal{O}(\p^3)\, ,
\eea
where we have absorbed the higher derivative terms into $\mathcal{O}(\p^3)$. The first term is a total divergence that fixes the number density and will not concern us here. The second term, however, is the Lorentz-invariant Klein-Gordon Lagrangian with ``speed of light'' given by the sound velocity $v$. What we have witnessed is the striking birth of {\it emergent} low energy Lorentz symmetry from the high energy Galilean symmetry. We can quickly read off the low energy linear dispersion of the Bogoliubov quasi-particles from the second term in Eq.\ (\ref{phonon}).

It is possible to fix the coefficient in front of $(\p_i\varphi)^2$ in (\ref{phonon}) directly from the action (\ref{Sboson}). We perform a {\it finite} $U_b(1)$ symmetry transformation of the action $S$ with parameter $\varphi(x,t)$ that varies smoothly in space and time. Examining the change in the Lagrangian, we find
\be\la{trick}
\mathcal L[e^{i\varphi}\phi] - \mathcal L[\phi] =\rho_B \left[i\p_0\varphi - \frac{1}{2m}(\p_i\varphi)^2\right]\,,
\ee
where we have ``integrated out" the absolute value $|\phi|^2$ ({\it i.e.} replaced it by the mean field value $\rho_B=\langle|\phi|^2\rangle$). Comparing with Eq.\ (\ref{phonon}), we see that Eq.\ (\ref{trick}) fixes the coefficients 
\be
\begin{array}{ccc}
\rho_0=\rho_B\ & \mathrm{and} & m^*=m,
\end{array}
\ee
but does not fix the sound velocity $v$.

\section{SUSY algebra}
Here we summarize the (super)algebra of SUSY transformations. 
Recall that we have as fundamental microscopic fields a complex scalar boson $\phi$ and a spinless complex fermion $\psi$ which satisfy
\begin{align}
\left[\phi(x),\phi(y)\right]&=\left\{\psi(x),\psi(y)\right\}=0, \\
\left[\phi(x),\phi^\dag(y)\right]&=\left\{\psi(x),\psi^\dag(y)\right\}=\delta(x-y), \\
\left[\phi(x),\psi(y)\right]&=\left[\phi^\dag(x),\psi(y)\right]=0.
\end{align}
In addition, we introduce Grassmann-valued supercharge operators given by
\begin{equation}
Q=\int d^dx \phi^\dag\psi = \int d^d x \mathcal{Q},\, \quad \bar{Q}=\int d^dx \psi^\dag\phi = \int d^d x \bar{\mathcal{Q}}\, ,
\end{equation}
together with two additional bosonic generators
\begin{align}
{N}&=\int d^dx \left(\phi^\dag\phi+\psi^\dag\psi\right)\, ,\\
{Y}&=\frac{1}{2}\int d^dx \left(\phi^\dag\phi-\psi^\dag\psi\right)\, .
\end{align}

These satisfy the super-commutation relations
\begin{align}
\left[Q,{N}\right]&=\left[\bar{Q},N\right]=\left[{Y},{N}\right]=0\label{alg1}\, ,\\
\left\{Q,Q\right\}&=\left\{\bar{Q},\bar{Q}\right\}=0\, ,\\
\left[Q,{Y}\right]&=-Q\, ,\\
\left[\bar{Q},{Y}\right]&=\bar{Q}\, ,\\
\left\{Q,\bar{Q}\right\}&={N}\label{alg2}\,.
\end{align}
Note that this is a supersymmetric generalization of the commutation relations of $\mathfrak{su}(2)$ - in fact, this is the superalgebra known as $\mathfrak{u}(1|1)$ \cite{Kac}. That is, $Y$ acts like the $z$-component of angular momentum, and $Q$ ($\bar{Q}$) acts like an angular momentum raising (lowering) operator. Furthermore, the number operator $N$ is the quadratic Casimir operator. The main difference lies in the fact that $Q$ and $\bar{Q}$ satisfy {\it anticommutation} relations. 

When acting on the elementary bose and fermi fields of our model, these operators satisfy
\begin{align}
\left[Q,\phi\right]&=-\psi\, ,\\
\left\{Q,\psi\right\}&=0\, ,\\
\left[\bar{Q},\phi\right]&=0\, ,\\
\left\{\bar{Q},\psi\right\}&=\phi\, ,\\
\left[N,\phi\right]&=-\phi\, ,\\
\left[N,\psi\right]&=-\psi\, ,\\
\left[Y,\phi\right]&=-\frac{1}{2}\phi\, ,\\
\left[Y,\psi\right]&=\frac{1}{2}\psi\, .
\end{align}
The particle exchange supersymmetry discussed in the main text is generated by the operator
\begin{equation}
{\tau}=\left(\bar{\eta} Q+\eta\bar{Q}\right)\, .
\end{equation}
Under this transformation, the elementary fields transform as
\begin{align}
\phi&\rightarrow\left(1+\frac{\eta\bar{\eta}}{2}\right)\phi-\bar{\eta}\psi\, ,\\
\psi&\rightarrow\left(1-\frac{\eta\bar{\eta}}{2}\right)\psi+\eta\phi\, .
\end{align}
This is an explicit form of the matrix transformation law mentioned in the main text.

For completeness, we note that this algebra admits a family of representations in terms of Pauli matrices:
\begin{align}
Q&\rightarrow\frac{\alpha}{2}\sigma_{+}\, , \\
\bar{Q}&\rightarrow\frac{\bar{\alpha}}{2}\sigma_{-}\, ,\\
Y&\rightarrow\frac{1}{2}\sigma_z\, ,\\
{N}&\rightarrow \left|\alpha\right|^2I\, .
\end{align}
{for any complex number $\alpha$}.

\section{Approximately broken symmetries and almost-gapless excitations}

Here we will review some field-theoretic arguments for the existence of gapped collective excitations when a symmetry is explicitly broken. In the process, we will uncover some justification for the mean-field ``trick" we used in the main text to fix the goldstino dispersion relation. Our discussion follows that of Weinberg's text \cite{Weinberg}, with only small modifications. 

We start by recalling that for a quantum system with action $S[\phi,\psi]$, we can form the generating functional (for brevity, we suppress the explicit dependence of functionals on the Hermitian conjugate of the variables. We hope this does not cause undue confusion)
\begin{equation}
e^{iW[J,\theta]}=\int\mathcal{D}\phi\mathcal{D}\psi e^{iS[\phi,\psi]+i\int{d^dx}J\phi+\theta\psi+\mathrm{h.c.}}.
\end{equation}
$W$ is the generating functional for connected correlation functions (cumulants) of the quantum system; that is, functional derivatives of $W$ with respect to the source fields $J$ and $\theta$ give - upon setting the sources to zero - the connected averages of products of the fields $\phi$ and $\psi$. From $W$, we can define the effective action $\Gamma$ via a Legendre transform:
\begin{equation}
\Gamma(\phi,\psi)=W-\int{d^dx}J\phi+\theta\psi+\mathrm{h.c.}.
\end{equation}
The effective action $\Gamma$ is the quantum analogue of the classical action - the averages of the external fields in the absence of sources are given by extremal points of $\Gamma$. 

The effective action has many other useful properties, although we will only need two of them here. First, the matrix of second variational derivatives of $\Gamma$ is precisely the negative of the inverse propagator. Using the superfield notation of Eq. (3) in the main text, this means
\begin{equation}
\frac{\delta^2\Gamma}{\delta\Phi_i(x)\delta\Phi^\dag_j(y)}=-\left<\Phi_i(x)\Phi^\dag_j(y)\right>^{-1}\equiv\Delta_{ij}^{-1}(x-y),
\end{equation}
when evaluated at the extrema of $\Gamma$. The second property we need is that linear symmetries of the microscopic action are automatically also symmetries of the effective action.

Since we are only interested in homogeneous configurations of our system, it is sufficient for us to consider the effective potential $\mathcal{V}$ defined by the effective action $\Gamma$ evaluated on uniform configurations of the external fields. 

Let us consider first the case when supersymmetry is an exact symmetry of the microscopic action, and ask what happens when it is spontaneously broken. Spontaneous breakdown of the symmetry means, in this language, that the minimum of the effective potential occurs at a nonzero value of at least one of the fields. The invariance of the microscopic action under supersymmetry implies for the effective potential that
\begin{equation}
0=\frac{\partial \mathcal{V}}{\partial\Phi_i}N_{ij}\Phi_j+\frac{\partial \mathcal{V}}{\partial\Phi^\dag_i}N^\dag_{ij}\Phi_j^\dag\, . \label{effpotsym}
\end{equation}
Taking an additional derivative with respect to $\Psi_j$ or $\Psi_j^\dag$ and evaluating at the \emph{nonzero} minimum of the effective potential tells us immediately that the inverse propagator
$\Delta_{ij}^{-1}$ has a zero (super-)determinant at vanishing momentum (since we are looking at homogeneous configurations). We conclude that the two-point function for the matter fields has a pole at zero wavevector - there is a gapless collective excitation. This generalizes Goldstone's theorem to the case of Grassmann-valued symmetries; it yields a fermionic collective mode at zero energy.

We would now like to ask what happens when supersymmetry is explicitly broken in the microscopic action. In this case, the RHS of Eq. (\ref{effpotsym}) will no longer be zero. In fact, carrying through the same derivation as above, the right hand side, however complicated it is, gives the chemical potnetial for the fermionic collective mode.

In general, the variation of the effective potnetial under a SUSY transformation is given by matrix elements of the supercurrent operator [Eq.(3) in the main text],
which can be calculated in perturbation theory. However, this is where our mean-field trick comes to the rescue. Considering the change in the microscopic action under spatially varying SUSY transformations is precisely equivalent to calculating the relevant matrix elements to tree level! Thus, our derivation of the fermionic sound spectrum in the main text is justified whenever perturbation theory is valid.

\bibliography{Bibliography}